\documentclass[%
reprint,
superscriptaddress,
amsmath,amssymb,
aps,
floatfix,
longbibliography
]{revtex4-1}

\usepackage{graphicx}
\usepackage{multirow}
\usepackage{dcolumn}
\usepackage{bm}
\usepackage[dvipsnames]{xcolor}

\begin{document}

\title{Magnetic and electronic phase transitions probed by nanomechanical resonance}

\author{Makars \v{S}i\v{s}kins}
\thanks{These authors contributed equally.}
\author{Martin Lee}%
\thanks{These authors contributed equally.}
\affiliation{%
 Kavli Institute of Nanoscience, Delft University of Technology, Lorentzweg 1,\\
 2628 CJ, Delft, The Netherlands
}

\author{Samuel Ma\~{n}as-Valero}%
\author{Eugenio Coronado }%
\affiliation{%
Instituto de Ciencia Molecular (ICMol), Universitat de Val\`{e}ncia, c/Catedr\'{a}tico Jos\'{e} Beltr\'{a}n 2,\\ 46980 Paterna, Spain
}%
\author{Yaroslav M. Blanter}%
\author{Herre S. J. van der Zant}
\affiliation{%
	Kavli Institute of Nanoscience, Delft University of Technology, Lorentzweg 1,\\
	2628 CJ, Delft, The Netherlands
}

\author{Peter G. Steeneken}
\email{e-mail: p.g.steeneken@tudelft.nl; h.s.j.vanderzant@tudelft.nl; m.siskins-1@tudelft.nl}
\affiliation{%
	Kavli Institute of Nanoscience, Delft University of Technology, Lorentzweg 1,\\
	2628 CJ, Delft, The Netherlands
}%
\affiliation{
	Department of Precision and Microsystems Engineering, Delft University of Technology,
	Mekelweg 2,\\ 2628 CD, Delft, The Netherlands}

\begin{abstract}
Two-dimensional (2D) materials enable new types of magnetic and electronic phases mediated by their reduced dimensionality like magic-angle induced phase transitions \cite{Cao2018,Sharpe2019}, 2D Ising antiferromagnets \cite{Lee2016} and ferromagnetism in 2D atomic layers \cite{Huang2017,Gong2017} and heterostructures \cite{Gibertini2019}. However, only a few methods are available to study these phase transitions \cite{Lee2016,McGuire2015,Huang2017,Novoselov2016,Gibertini2019}, which for example is particularly challenging for antiferromagnetic materials \cite{Gibertini2019}. Here, we demonstrate that these phases can be probed by the mechanical motion: the temperature dependent resonance frequency and quality factor of multilayer 2D material membranes show clear anomalies near the phase transition temperature, which are correlated to anomalies in the specific heat of the materials. The observed coupling of mechanical degrees of freedom to magnetic and electronic order is attributed to thermodynamic relations that are not restricted to van der Waals materials. Nanomechanical resonators, therefore, offer the potential to characterize phase transitions in a wide variety of materials, including those that are antiferromagnetic, insulating or so thin that conventional bulk characterization methods become unsuitable.
\end{abstract} 

\maketitle

A universal method to characterize phase transitions in bulk crystals is via anomalies in the specific heat, that are present at the transition temperature according to Landau's theory \cite{Landau1984}.  However, specific heat is difficult to measure in thin micron-sized samples with a mass of less than a picogram \cite{Morell2019,Dolleman2018}. Although coupling between mechanical and electronic/magnetic degrees of freedom might not seem obvious, the intuitive picture behind this coupling is that changes in the electronic/magnetic order and entropy in a material are reflected in its specific heat, which in turn results in variations in the thermal expansion coefficient that affect the tension and resonance frequency. As the specific heat near a phase transition is expected to exhibit a discontinuity \cite{Landau1984}, the temperature dependent resonance frequency of a suspended membrane can thus be used to probe this transition. Here, we use nanomechanical motion to investigate magnetic order in membranes of semiconducting FePS$_3$, NiPS$_3$ and insulating MnPS$_3$ - antiferromagnetic members of the transition-metal phosphor trisulphides (MPS$_3$) \cite{Joy1992}, and subsequently discuss results on metallic 2H-TaS$_2$, which exhibits a transition to a charge density wave state \cite{AbdelHafiez2016}.

\begin{figure*}[ht]
	\includegraphics[scale=0.50]{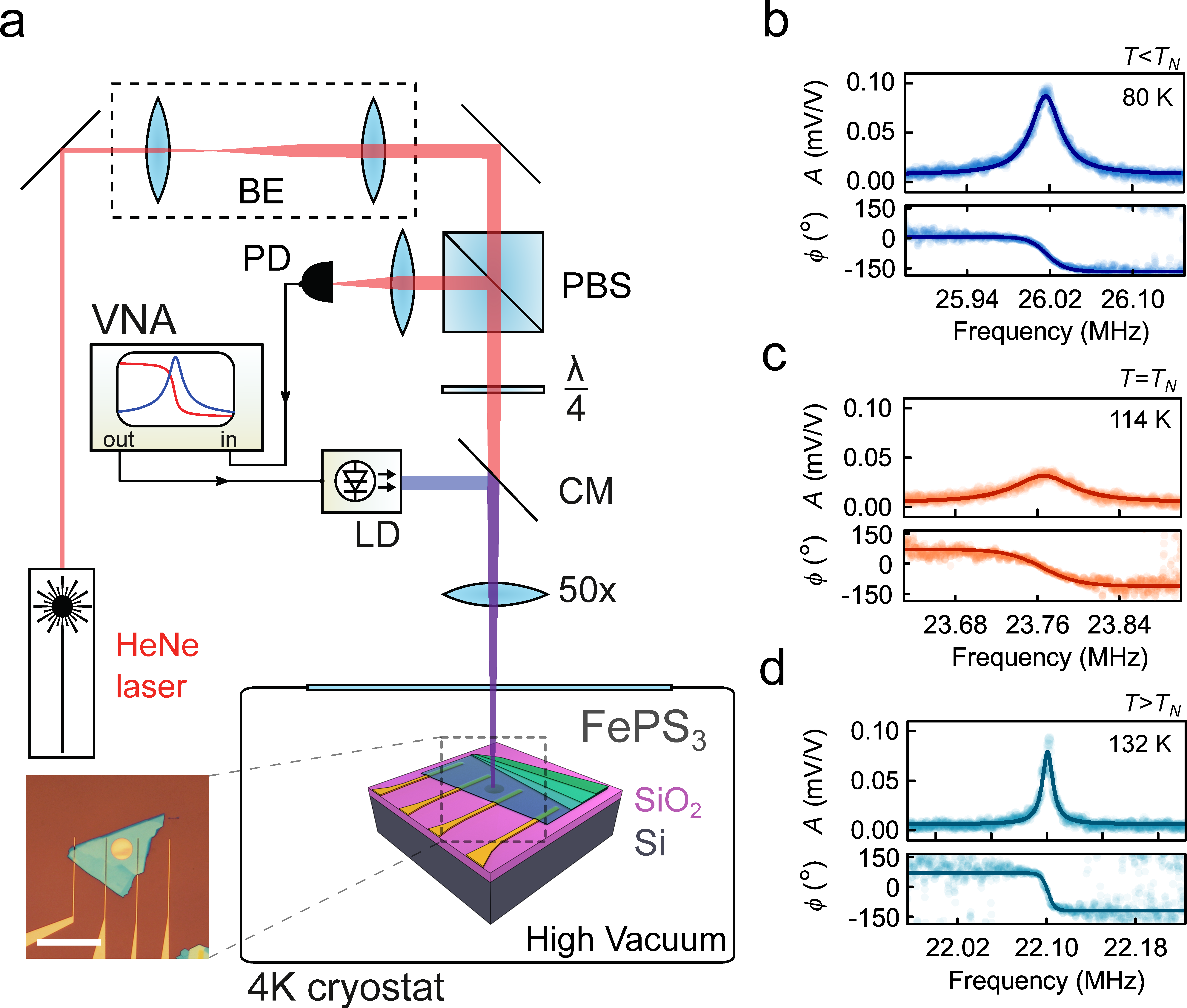}
	\caption{Characterisation of mechanical resonances in a thin antiferromagnetic FePS$_3$ membrane. (a) Laser interferometry setup. Red interferometric detection laser:  $ \lambda_{\mathrm{red}} = 632 $ nm. Blue actuation laser diode: $\lambda_{\mathrm{blue}} = 405$ nm. VNA - Vector Network Analyzer, CM - Cold Mirror, PBS - Polarizing Beam Splitter, PD - Photodiode, LD - Laser Diode. Inset: optical image of a FePS$_3$ membrane, including electrodes introducing an option for electrostatic control of strain in the membrane. Flake thickness: $45.2\pm0.6$ nm; membrane diameter: $d=10$ $\mu$m.  Scale bar: $30$ $\mu$m. (b-d) Amplitude ($A$) and phase ($\phi$) of the fundamental resonance at three different temperatures for the device shown in (a). Filled dots - measured data; solid lines - fit of the mechanical resonance used to determine $f_0$ and $Q$ \cite{CastellanosGomez2013}.}
	\label{fgr:first}
\end{figure*}

FePS$_3$ is an Ising-type antiferromagnet with a N\`{e}el temperature in bulk in the range of $T_N\sim118-123$ K \cite{Lee2016,Joy1992,Takano2004}, exhibiting a distinct feature in its specific heat near $T_N$ \cite{Takano2004}. Ionic layers in FePS$_3$ are stacked in van der Waals planes, that can be exfoliated to thin the crystal down with atomic precision \cite{Lee2016}. Using mechanical exfoliation and all-dry viscoelastic stamping \cite{CastellanosGomez2014}, we transfer thin flakes of FePS$_3$ over circular cavities etched in an oxidised Si wafer, to form membranes (see the inset in Fig.~\ref{fgr:first}a). Suspended FePS$_3$ devices with thicknesses ranging from $8$ to $45$ nm are placed in a cryostat and cooled down to a temperature of $4$ K. The resonance frequency of the nanodrums is then characterized using a laser interferometry technique \cite{CastellanosGomez2013} (see Fig.~\ref{fgr:first}a and Methods).

\begin{figure*}[ht]
	\includegraphics[scale=0.56]{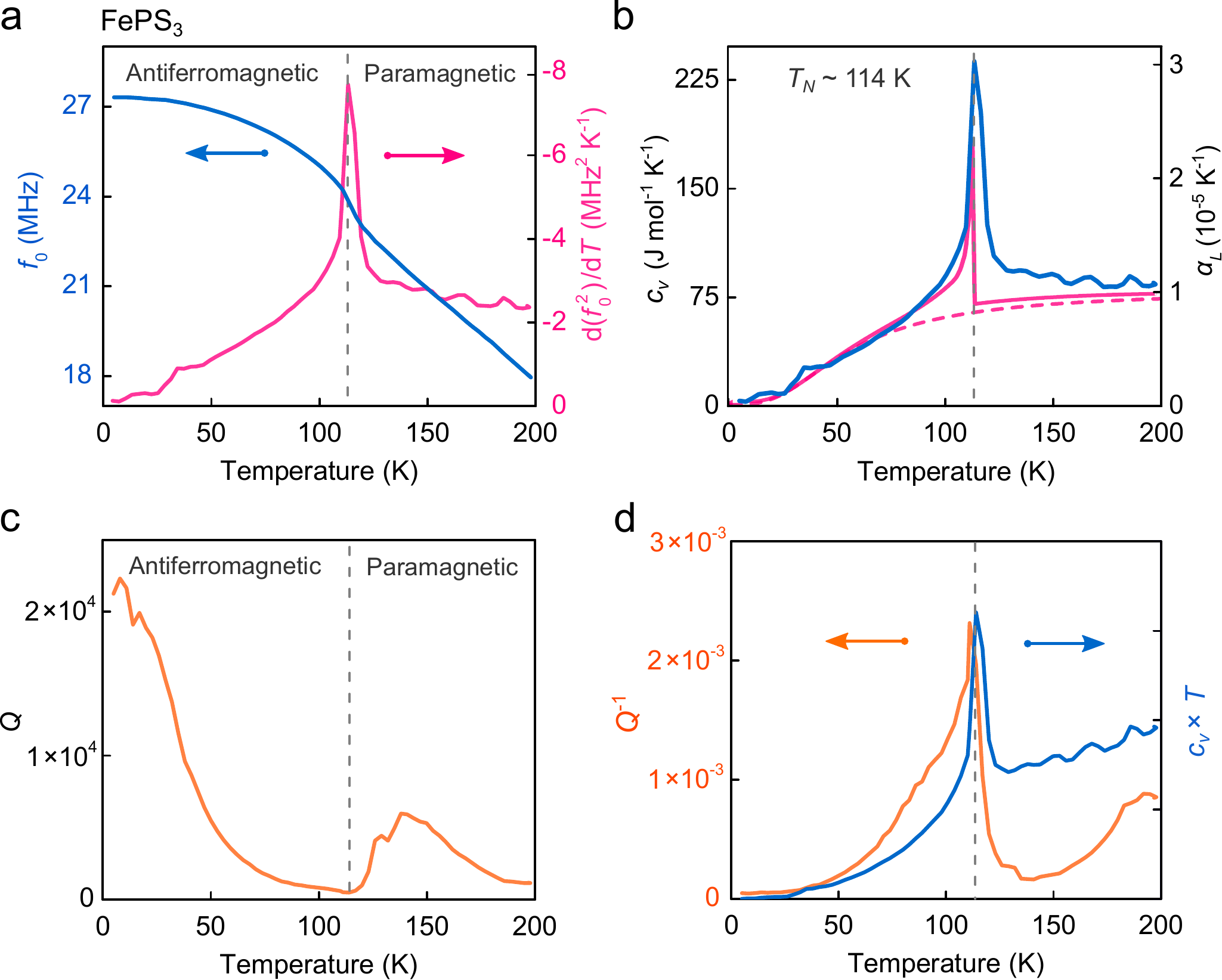}
	\caption{Mechanical and thermal properties properties of a FePS$_3$ resonator with membrane thickness of $45.2\pm0.6$ nm. In all panels, dashed vertical lines indicate the detected transition temperature, $T_N=114\pm3$ K as determined from the peak in the temperature derivative of $f^2_0$. (a) Solid blue line - measured resonance frequency as a function of temperature. Solid magenta line - temperature derivative of $f^2_0$. (b) Solid blue line - experimentally derived specific heat and corresponding thermal expansion coefficient. Solid magenta line - the theoretical calculation of the magnetic specific heat as reported in Takano, \textit{et al.} \cite{Takano2004} added to the phononic specific heat from Debye model (dashed magenta line) with a Debye temperature of $\Theta_D=236$ K \cite{Takano2004}. (c) Mechanical quality factor $Q(T)$ of the membrane fundamental resonance. (d) Solid orange line - measured mechanical damping $Q^{-1}(T)$ as a function temperature. Solid blue line - normalized $c_v(T)\, T$ term \cite{Zener1937, Lifshitz2000} (see Supplementary Section 4), with $c_v(T)$ taken from (b).}
	\label{fgr:magnetic}
\end{figure*}

The resonance frequency of the fundamental membrane mode, $f_0(T)$, is measured in the temperature range from $4$ to $200$ K. Typical resonances are shown in Fig.~\ref{fgr:first}b-d in the antiferromagnetic phase ($80$ K), near the transition ($114$ K) and in the paramagnetic phase ($132$ K), respectively. Figure \ref{fgr:magnetic}a shows $f_0(T)$ of the same FePS$_3$ membrane (solid blue curve). Near the phase transition, significant changes in amplitude, resonance frequency and quality factor are observed. 
To analyze the data further, we first analyze the relation between $f_0$ and the specific heat. The decrease in resonance frequency with increasing temperature in Fig.~\ref{fgr:magnetic}a is indicative of a reduction in strain due to thermal expansion of the membrane. The observed changes can be understood by considering the resonance frequency of a bi-axially tensile strained circular membrane:
\begin{equation}\label{eq:freq}
f_0(T)=\frac{2.4048}{\pi d}\sqrt{\frac{E}{\rho}\frac{\epsilon(T)}{(1-\nu)}},
\end{equation}
where $E$ is the Young's modulus of the material, $\nu$ its Poisson's ratio, $\rho$ its mass density, $\epsilon(T)$ the strain and $T$ the temperature. The linear thermal expansion coefficient of the membrane, $\alpha_L(T)$, and silicon substrate, $\alpha_{\rm Si}(T)$, are related to the strain in the membrane \cite{Morell2016,Singh2010} as $\frac{\mathrm{d}\epsilon(T)}{\mathrm{d}T}\approx-(\alpha_L(T)-\alpha_{\mathrm{Si}}(T))$, using the approximation $\alpha_{\mathrm{SiO_2}} \ll \alpha_{\mathrm{Si}}$ (see Supplementary Section 1). By combining the given expression for $\frac{\mathrm{d}\epsilon(T)}{\mathrm{d}T}$ with equation (\ref{eq:freq}) and by using the thermodynamic relation $\alpha_L(T)= \gamma c_v(T)/(3KV_M)$ \cite{Sanditov2011} between $\alpha_L(T)$ and the specific heat (molar heat capacity) at constant volume, $c_v(T)$, we obtain:
\begin{equation}\label{eq:heatrelation}
c_v(T) = 3\alpha_L(T)\frac{K V_M}{\gamma}  = 3\left(\alpha_{\rm Si} - \frac{1}{\mu^2}\frac{{\rm d} [f_0^2(T)]}{{\rm d}T}\right)\frac{K V_M}{\gamma}.
\end{equation}
Here, $K$ is the bulk modulus, $\gamma$ the Gr\"{u}neisen parameter, $V_M=M/\rho$ the molar volume of the membrane and $\mu=\frac{2.4048}{\pi d}\sqrt{\frac{E}{\rho(1-\nu)}}$, that are assumed to be only weakly temperature dependent. The small effect of non-constant volume ($\nu \ne 0.5$) on $c_v$ is neglected. 

We use the equation (\ref{eq:heatrelation}) to analyze $f_0(T)$ and compare it to the calculated specific heat for FePS$_3$ from literature \cite{Takano2004}. In doing so, we estimate the Gr\"{u}neisen parameter following the Belomestnykh$-$Tesleva relation $\gamma\approx\frac{3}{2}\left(\frac{1+\nu}{2-3\nu}\right)$ \cite{Belomestnykh2004,Sanditov2011}. This is an approximation to Leont'ev's formula \cite{Leontiev1981}, which is a good estimation of $\gamma$ for bulk isotropic crystalline solids within $\sim10\%$ of uncertainty \cite{Sanditov2011}. Furthermore, we use literature values for the elastic parameters of FePS$_3$ as obtained from first-principles theoretical calculations \cite{Hashemi2017} to derive $E=103$ GPa, $\nu=0.304$ and $\rho=3375$ kg/m$^3$ (see Supplementary Section 2). 
In Fig.~\ref{fgr:magnetic}a, the steepest part of the negative slope of $f_0(T)$ (solid blue curve) leads to a large peak in $\frac{{\rm d}(f_0^2(T))}{{\rm d}T}$ (solid magenta curve) near $114$ K, the temperature which we define as $T_N$ and indicate by the vertical dashed lines. In Fig.~\ref{fgr:magnetic}b the specific heat curve of FePS$_3$ (blue solid line) as estimated from the data in Fig.~\ref{fgr:magnetic}a and equation (\ref{eq:heatrelation}) is displayed. The results are compared to a theoretical model for the specific heat of FePS$_3$ (magenta solid line in Fig.~\ref{fgr:magnetic}b), which is the sum of a phononic contribution from the Debye model (magenta dashed line) and a magnetic contribution as calculated by Takano, \textit{et al.} \cite{Takano2004}. It is noted that other, e.g. electronic contributions to $c_v(T)$ are small and can be neglected in this comparison, as is supported by experiments on the specific heat in bulk FePS$_3$ crystals \cite{Takano2004}. The close correspondence in Fig.~\ref{fgr:magnetic}b between the experimental and theoretical data for $c_v(T)$ supports the applicability of equation (\ref{eq:heatrelation}). It also indicates that changes in the Young’s modulus near the phase transition, that can be of the order of a couple of percent \cite{Barmatz1975}, are insignificant and that it is the anomaly in $c_v$ of FePS$_3$ which produces the observed changes in resonance frequency and the large peak in $\frac{{\rm d}(f_0^2)}{{\rm d}T}$ visible in Fig.~\ref{fgr:magnetic}a.

The abrupt change in $c_v(T)$ of the membrane can be understood from Landau's theory of phase transitions \cite{Landau1984}. To illustrate this, we consider a simplified model for an antiferromagnetic system, like FePS$_3$, with free energy, $F$, which includes a strain-dependent magnetostriction contribution (see Supplementary Section 3). Near the transition temperature and in the absence of a magnetic field it holds that:
\begin{equation}\label{eq:freeenergy}
    F = F_0 + [a(T-T_N)+\zeta(\epsilon)]L_z^2+BL_z^4.
\end{equation}
Here, $a$ and $B$ are phenomenological positive constants, $L_z$ is the order parameter in the out-of-plane direction and $\zeta (\epsilon)=\eta_{ij} \epsilon_{ij}$, a strain-dependent parameter with $\eta_{ij} $ a material-dependent tensor, that includes the strain and distance dependent magnetic exchange interactions between neighbouring magnetic moments. By minimizing equation (\ref{eq:freeenergy}) with respect to $L_z$, the equilibrium free energy, $F_{min}$, and order parameter are obtained (see Supplementary Section 3). Two important observations can be made. Firstly, strain shifts the transition temperature according to:
\begin{equation}\label{eq:Tn}
    T_N^*(\epsilon) = T_N - \frac{\zeta (\epsilon)}{a},
\end{equation} 
where $T_N^*$ is the Ne\'el temperature, below which free energy minima $F_{min}$ with finite order ($L_z \ne 0$) appear. Secondly, since close to the transition the specific heat follows $c_{v}(T)=-T\frac{\partial^2F_{min}}{\partial T^2} $, this general model predicts a discontinuity in $c_v$ of magnitude $T_N^{*}\frac{a^2}{2B}$ at the transition temperature $T_N^*$, in accordance with the experimental jump in $c_v(T)$ and $\frac{{\rm d}(f_0^2(T))}{{\rm d}T}$ observed in Fig.~\ref{fgr:magnetic}a and b.
\begin{figure*}[ht]
	\includegraphics[scale=0.55]{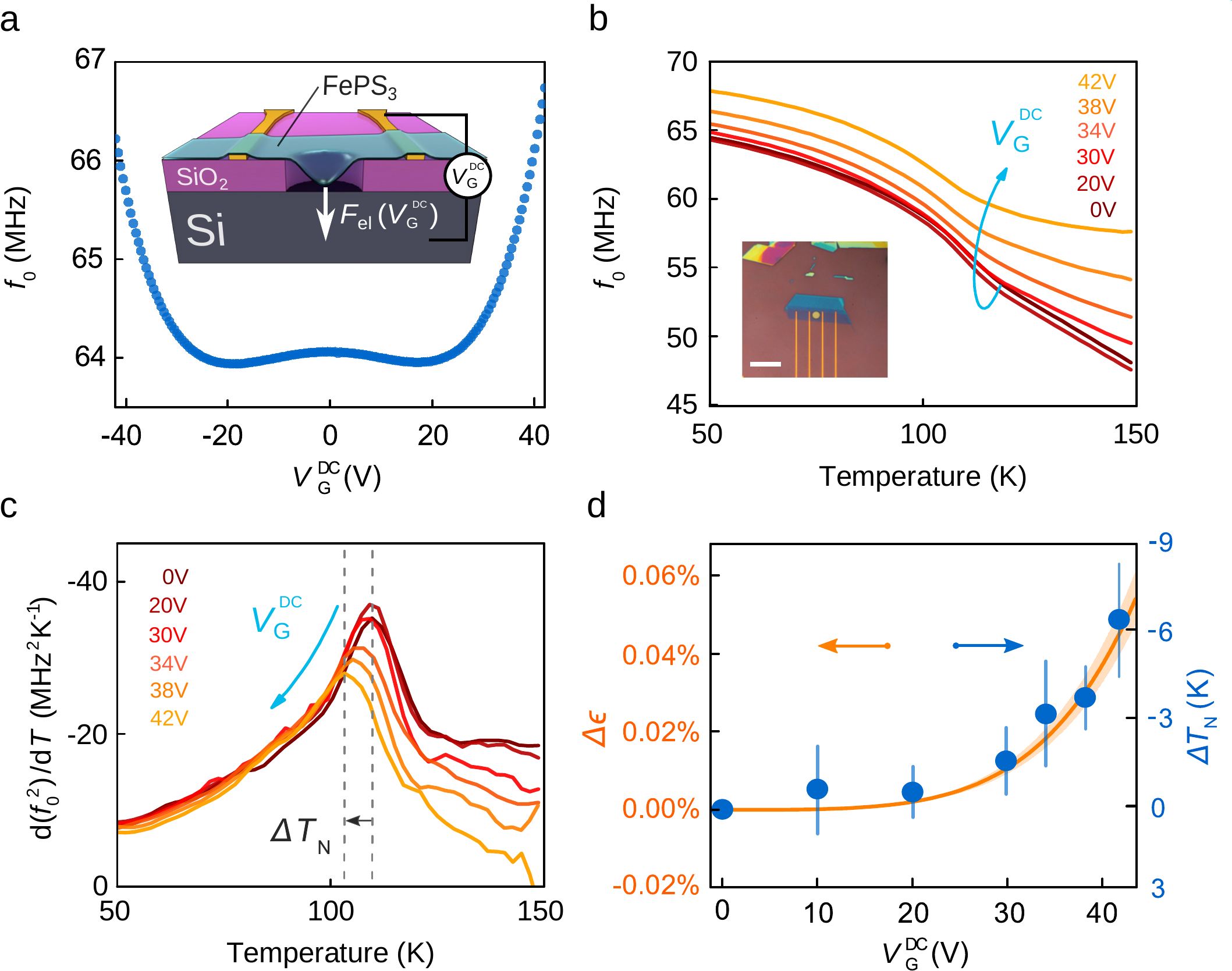}
	\caption{Resonance frequency and transition temperature tuning with a gate voltage. (a) Resonance frequency as a function of gate voltage at $50$ K. Inset - schematics of the electrostatic tuning principle. (b) Resonance frequency as a function of temperature for six different voltages. Inset: optical image of the sample, $t=8\pm0.5$ nm. Scale bar: $16$ $\mu$m. (c) Derivative of $f_0^2$ as a function of gate voltage and temperature. Blue arrow, line colors and legend indicate the values of $V_G^{\mathrm{DC}}$. Dashed grey lines indicate the decrease in transition temperature $\Delta T_N=T^*_N(V_G^{\mathrm{DC}})-T_N(0\: \mathrm{V})$ with increasing $V_G^{\mathrm{DC}}$. (d) Blue solid dots - shift in $T_N$ as a function of $V_G^{\mathrm{DC}}$. Orange solid line - model of electrostatically induced strain $\Delta \epsilon$ as a function of $V_G^{\mathrm{DC}}$ (see Supplementary Section 5).}
	\label{fgr:pulling}
\end{figure*}

We now analyze the quality factor data shown in Fig.~\ref{fgr:magnetic}c,d. Just above $T_N$, the quality factor of the resonance (Fig.~\ref{fgr:magnetic}c) shows a significant increase as the temperature is increased from $114$ to $140$ K. The observed minimum in the quality factor near the phase transition, suggests that dissipation in the material is linked to the thermodynamics and can be related to thermoelastic damping. We model the thermoelastic damping according to Zener \cite{Zener1937} and Lifshitz-Roukes \cite{Lifshitz2000} that report dissipation of the form $Q^{-1} =\beta c_v(T)\, T$, where $\beta$ is the thermomechanical term (see Supplementary Section 4). Since we have obtained an estimate of $c_v(T)$ from the resonance frequency analysis (Fig.~\ref{fgr:magnetic}b), we use this relation to compare the experimental dissipation $Q^{-1}(T)$ (orange solid line) to a curve proportional to $c_v(T) \, T$ (blue solid line) in Fig.~\ref{fgr:magnetic}d. Both the measured dissipation and the thermoelastic term display a peak near $T_N\sim114$ K. The close qualitative correspondence between the two quantities is an indication that the thermoelastic damping related term indeed can account for the temperature dependence of $Q(T)$ near the phase transition. We note that the temperature dependent dissipation in thin membranes is still not well understood, and that more intricate effects might play a role in the observed temperature dependence.\begin{figure}[ht]
	\includegraphics[scale=0.39]{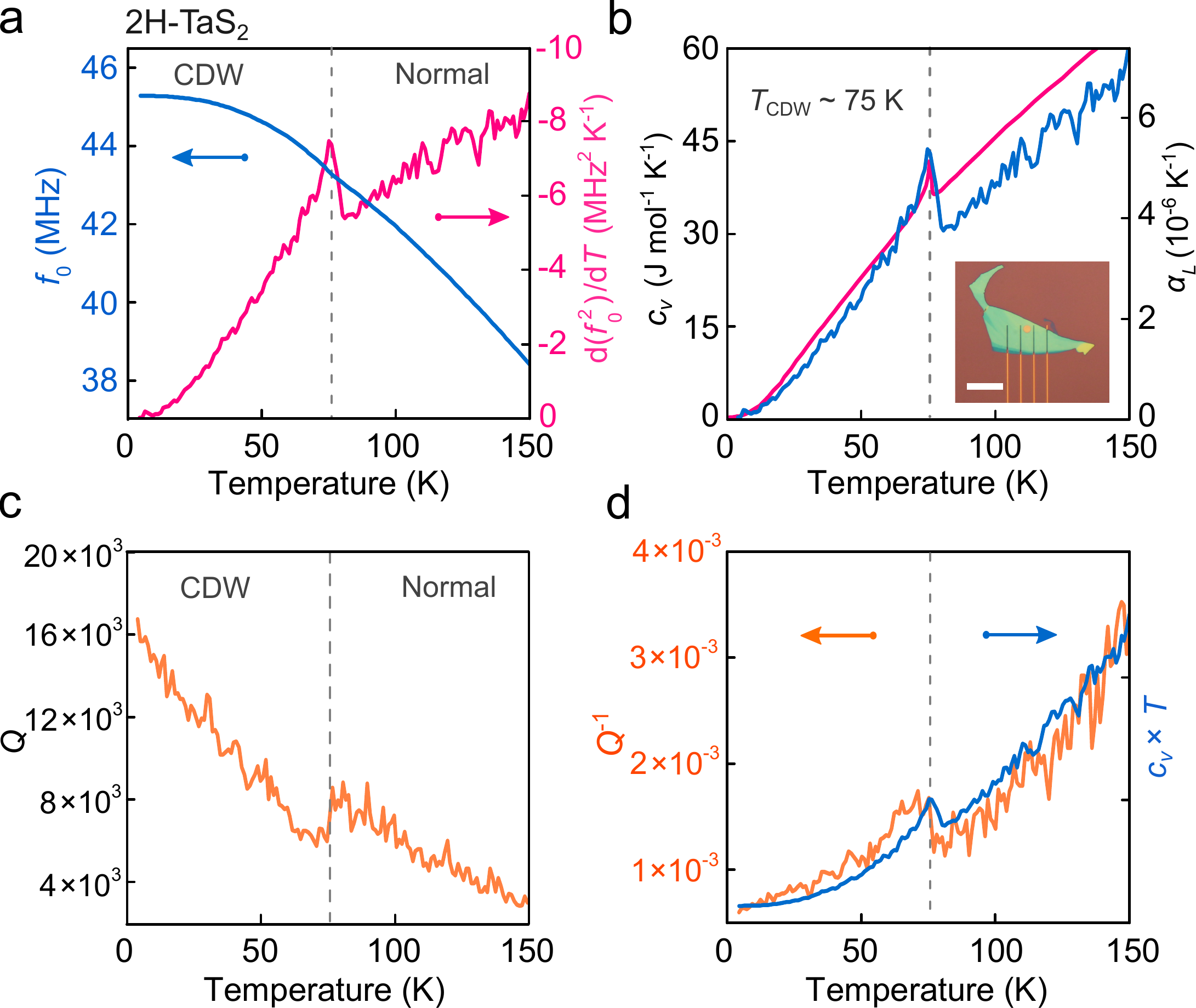}
	\caption{Mechanical properties of a 2H-TaS$_2$ resonator with membrane thickness of $31.2\pm0.6$ nm and $d=4$ $\mu$m. Dashed vertical line in all $4$ panels indicates the detected $T_{\mathrm{CDW}}$, that is defined by the peak in $\frac{{\rm d}(f_0^2(T))}{{\rm d}T}$. (a) Solid blue line - resonance frequency as a function of temperature. Solid magenta line - temperature derivative of $f^2_0$. (b) Solid blue line - experimentally derived $c_v$ and thermal expansion coefficient as a function of temperature (see Supplementary section 6). Solid magenta line - specific heat of bulk 2H-TaS$_2$ as reported in Abdel-Hafiez, \textit{et al.} \cite{AbdelHafiez2016}. Inset: optical image of the sample. Scale bar: $20$ $\mu$m. (c) Quality factor $Q(T)$ as a function of temperature. (d) Solid orange line - measured mechanical damping $Q^{-1}(T)$ as a function of temperature. Solid blue line - curve proportional to the term $c_v(T)\, T$ \cite{Zener1937, Lifshitz2000} (see Supplementary Section 4), with $c_v(T)$ taken from the experimental data in (b).}
	\label{fgr:Tas2}
\end{figure}

Equation (\ref{eq:Tn}) predicts that the transition temperature is strain-dependent due to the distance dependent interaction coefficient $\zeta(\epsilon)$ between magnetic moments. To verify this effect, we use an $8\pm0.5$ nm thin sample of FePS$_3$ suspended over a cavity of $4$ $\mu$m in diameter. A gate voltage $V_G^{\mathrm{DC}}$ is applied between the flake and the doped bottom Si substrate to introduce an electrostatic force that pulls the membrane down and thus strains it (see Supplementary Section 5). As shown in Fig.~\ref{fgr:pulling}a, the resonance frequency of the membrane follows a W-shaped curve as a function of gate voltage. This is due to two counteracting effects \cite{Lee2018}: at small gate voltages capacitive softening of the membrane occurs, while at higher voltages the membrane tension increases due to the applied electrostatic force, which causes the resonance frequency to increase. 

Figure \ref{fgr:pulling}b shows $f_0(T)$ for six different gate voltages. The shift of the point of steepest slope of $f_0(T)$ with increasing $V_G^{\mathrm{DC}}$ is well visible in Fig.~\ref{fgr:pulling}b and even more clear in Fig.~\ref{fgr:pulling}c, where the peak in $\frac{{\rm d}(f_0^2)}{{\rm d}T}$ shifts $6$ K downward by electrostatic force induced strain. The observed reduction in $T_N^*$ as determined by the peak position in $\frac{{\rm d}(f_0^2)}{{\rm d}T}$ qualitatively agrees with the presented model and its strain dependence from equation (\ref{eq:Tn}), as shown in Fig.~\ref{fgr:pulling}d indicative of a reduced coupling of magnetic moments with increasing distance between them due to tensile strain.

Since the coupling between specific heat and the order parameter in materials is of a general thermodynamic nature, the presented methodology is applicable to a wide variety of materials provided that elastic properties of the material and Gr\"{u}neisen parameter are weakly temperature dependent, the substrate satisfies the condition $\alpha_{\mathrm{substrate}}\ll \alpha_{\mathrm{material}}$ and that the frequency shifts and changes in $Q$ are large enough to be resolved. We further demonstrate the method by detecting magnetic phase transitions in NiPS$_3$ and MnPS$_3$. Compared to FePS$_3$, the effect of the phase transitions in MnPS$_3$ and NiPS$_3$ on the resonances is more gradual (see Supplementary Section 2) with both materials showing broader maxima in $\frac{{\rm d}(f_0^2(T))}{{\rm d}T}$ near their $T_N$ at $76$ K and $151$ K, respectively, which is consistent with measurements of bulk crystals \cite{Takano2004,Joy1992}. 

In order to demonstrate the detection of an electronic phase transition, we now discuss results for 2H-TaS$_2$ that in bulk exhibits a charge density wave (CDW) transition at $T_{CDW}\sim77$ K \cite{AbdelHafiez2016}. Figure \ref{fgr:Tas2}a shows a transition-related anomaly in both $f_0(T)$ (solid blue line) and the temperature derivative of $f_0^2(T)$ (solid magenta line) that peaks at $75\pm3$ K. We convert $\frac{{\rm d}(f_0^2(T))}{{\rm d}T}$ to the corresponding $c_v(T)$ using the same approach as discussed before (see Supplementary Section 6). Figure \ref{fgr:Tas2}b shows a downward step in the specific heat at $75$ K (solid blue line), indicative of a phase transition from the CDW to the disordered high-temperature state \cite{Landau1984, SaintPaul2019} with a close quantitative correspondence to $c_v$ measured in a bulk crystal \cite{AbdelHafiez2016} (drawn magenta line). This anomaly occurs near the electrically determined phase transition temperature of $\sim77$ K on the same flake (see Supplementary Section 6) and is also consistent with the CDW transition temperature previously reported in 2H-TaS$_2$ \cite{AbdelHafiez2016}. The Q-factor also shows a local minimum with a drop next to the transition temperature (see Fig.~\ref{fgr:Tas2}c). As discussed before \cite{Zener1937, Lifshitz2000}, $Q^{-1}(T)$ is expected to follow the same trend as $c_v(T) \, T$. Both quantities are displayed in Fig.~\ref{fgr:Tas2}d and indeed show a good qualitative correspondence.

In conclusion, we have demonstrated a method for identifying phase transitions in ultrathin membranes of 2D materials via their mechanical resonance. An analytical equation for the relation between the specific heat of the material and the temperature dependent resonance frequency is derived and shown to be in good agreement with experimental results. The presented methodology thus shows that mechanical motion of suspended membranes can be used to probe magnetic and electronic order in membranes. Since the materials are characterised in a suspended state, substrate effects on the electronic and magnetic properties of the thin materials are excluded. The technique is particularly appealing for the characterisation of ultrathin membranes of antiferromagnetic and insulating materials that are difficult to characterize otherwise. It is anticipated that it can be applied to a large range of van der Waals materials \cite{Novoselov2016,Gibertini2019}, thin 2D complex oxide sheets \cite{Ji2019,davidovikj2019ultrathin} and organic antiferromagnets \cite{LpezCabrelles2018}, contributing to a better understanding of fundamental models of magnetism and other ordering mechanisms in two dimensions. 

\section*{Methods}
\paragraph*{Sample fabrication}
To realize electrical contact to the samples for electrostatic experiments, Ti/Au electrodes are pre-patterned by a lift-off technique. Cavities are defined by reactive ion etching of circular holes with a diameter of $4-10$ $\mu$m in oxidized doped silicon wafers with an SiO$_2$ thickness of $285$ nm. Flakes of van der Waals crystals are exfoliated from high quality synthetically grown crystals with known stoichiometry (see Supplementary Section 7). All flakes are transferred on a pre-patterned chip by an all-dry viscoelastic stamping directly after exfoliation. Subsequently, samples are kept in an oxygen free environment to avoid degradation.
\paragraph*{Controlled measurement environment}
The samples are mounted on a piezo-based $xy$ nanopositioning stage inside a chamber of a closed-cycle cryostat with optical access. A closed feedback loop controlled local sample heater is used to perform temperature sweeps at a rate of $\sim5$ K/min, while keeping the pressure in the chamber below 10$^{-6}$ mbar. During the data acquisition temperature is kept constant with $\sim10$ mK stability.
\paragraph*{Laser interferometry}
A blue diode laser ($\lambda = 405$ nm), which is power-modulated by a Vector Network Analyzer (VNA), is used to excite the membrane and optothermally drive it into motion. Displacements are detected by focusing a red He-Ne laser beam ($\lambda=632$ nm) on the cavity formed by the membrane and Si substrate. The reflected light, which is modulated by the position-dependent membrane motion, is recorded by a photodiode and processed by a phase-sensitive VNA. All measurements are performed at incident laser powers of $P_{\mathrm{red}}<10$ $\mu$W and $P_{\mathrm{blue}}<0.6$ $\mu$W. It is checked for all membranes that the resonance frequency changes due to laser heating are insignificant. Laser spot size is on the order of $\sim1$ $\mu$m. The uncertainty in measured transition temperatures is estimated from determining the peak position in $-\frac{{\rm d}(f_0^2(T))}{{\rm d}T}$ within $2\%$ accuracy in the measured maximum.
\paragraph*{Atomic Force Microscopy}
AFM inspections to determine sample thickness are performed in tapping mode on a Bruker Dimension FastScan AFM. We use cantilevers with spring constants of $k = 30-40$ N/m. Error bars on reported thickness values are determined by measuring three to five profile scans of the same flake.

\begin{acknowledgements}
M.\v{S}., M.L., H.S.J.v.d.Z. and P.G.S. acknowledge funding from the European Union's Horizon $2020$ research and innovation program under grant agreement number $785219$. H.S.J.v.d.Z., E.C. and S.M.-V. thank COST Action MOLSPIN CA$15128$; E.C. and S.M.-V. thank ERC AdG Mol-2D $788222$, the Spanish MINECO (Project MAT$2017$-$89993$-R co-financed by FEDER and the Unit of Excellence “Maria de Maeztu” MDM-$2015$-$0538$) and the Generalitat Valenciana (Prometeo Programme).
\end{acknowledgements}

\section*{Author contributions}
M.\v{S}., M.L., E.C., H.S.J.v.d.Z. and P.G.S. conceived the experiments. M.\v{S}. performed the laser interferometry measurements. M.L. fabricated and inspected the samples. S.M.-V. and E.C. synthesized and characterized the FePS$_3$, MnPS$_3$, NiPS$_3$ and 2H-TaS$_2$ crystals. M.\v{S}., Y.M.B., and P.G.S. analysed and modeled the experimental data. H.S.J.v.d.Z. and P.G.S. supervised the project. The manuscript was jointly written by all authors with a main contribution from M.\v{S}. All authors discussed the results and commented on the manuscript.

\include{supinfo.tex}

\end{document}


\title{SUPPLEMENTARY INFORMATION: Magnetic and electronic phase transitions probed by nanomechanical resonance}

\author{Makars \v{S}i\v{s}kins}
\thanks{These authors contributed equally.}
\author{Martin Lee}%
\thanks{These authors contributed equally.}
\affiliation{%
 Kavli Institute of Nanoscience, Delft University of Technology, Lorentzweg 1,\\
 2628 CJ, Delft, The Netherlands
}

\author{Samuel Ma\~{n}as-Valero}%
\author{Eugenio Coronado }%
\affiliation{%
Instituto de Ciencia Molecular (ICMol), Universitat de Val\`{e}ncia, c/Catedr\'{a}tico Jos\'{e} Beltr\'{a}n 2,\\ 46980 Paterna, Spain
}%
\author{Yaroslav M. Blanter}%
\author{Herre S. J. van der Zant}%
\affiliation{%
	Kavli Institute of Nanoscience, Delft University of Technology, Lorentzweg 1,\\
	2628 CJ, Delft, The Netherlands
}

\author{Peter G. Steeneken}
\email{e-mail: p.g.steeneken@tudelft.nl; h.s.j.vanderzant@tudelft.nl; m.siskins-1@tudelft.nl}
\affiliation{%
	Kavli Institute of Nanoscience, Delft University of Technology, Lorentzweg 1,\\
	2628 CJ, Delft, The Netherlands
}%
\affiliation{
	Department of Precision and Microsystems Engineering, Delft University of Technology,
	Mekelweg 2,\\ 2628 CD, Delft, The Netherlands}

\maketitle
\onecolumngrid
\tableofcontents
\bigskip
\bigskip

\section{Fundamental resonance frequency of a circular plate and membrane}
\begin{figure*}[ht]
	\includegraphics[scale=0.25]{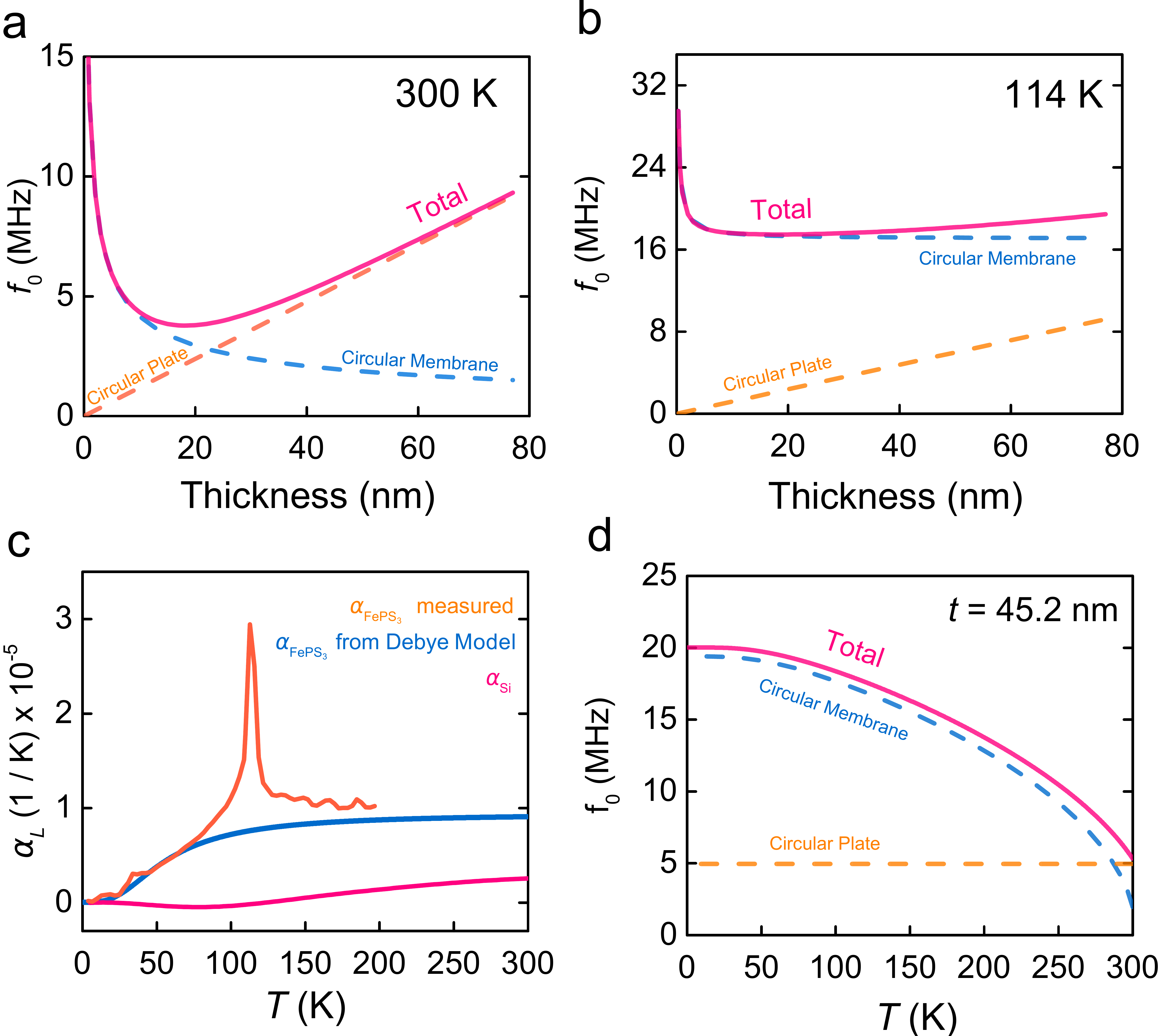}
	\caption{Resonance frequency of a FePS$_3$ resonator as a function of thickness for (a) $300$ K and (b) $114$ K. (c) Comparison of thermal expansion coefficients of FePS$_3$ as measured (solid orange line) and as predicted from the Debye model (solid blue line) to that of the Si substrate (solid magenta line). (d) Calculated resonance frequency of a FePS$_3$ membrane using equation (\ref{eq:SI_freq}) with a diameter of $10$ $\mu$m taking into account the thermal expansion coefficient of FePS$_3$ as predicted by the Debye model.}
	\label{fgr:plate-membrane}
\end{figure*}

In this section we analyze in more detail the resonance frequency of the FePS$_3$ resonators and show that near the phase transition they are close to the membrane limit. The fundamental resonance frequency of the mechanical resonator, $f_0$, in the crossover membrane-plate regime can be approximated as \cite{Wah1962,CastellanosGomez2013}:
\begin{equation}\label{eq:SI_freq}
\begin{split}
f_0&(T)\approx \sqrt{f^2_{membrane}+f^2_{plate}} =\\
   &\sqrt{\left(\frac{2.4048}{\pi d}\right)^2\frac{E}{\rho}\frac{\epsilon(T)}{(1-\nu)}+\left(\frac{10.21t}{\pi d^2}\right)^2\frac{E}{3\rho\left(1-\nu^2\right)}},
\end{split}
\end{equation}
where $d$ is the diameter of the membrane, $E$ the Young's modulus, $\epsilon (T)$ the strain, $\nu$ the Poisson's ratio, $t$ the thickness, $\rho$ the mass density and $T$ the temperature. The resonance frequency of the fundamental mode of a circular resonator is thickness dependent. For plate resonators $f_{plate} \propto t$, as expected for small and linear deflection \cite{Wah1962,CastellanosGomez2013, timoshenko1974theory}. For membranes, however, $f_0$ is dominated by the biaxial tension $N$:
\begin{equation}\label{eq:freq_memb}
f_{membrane}=\frac{2.4048}{\pi d}\sqrt{\frac{N}{\rho t}}.
\end{equation}
Equation (\ref{eq:freq_memb}) yields $f_{membrane} \propto t^{-0.5}$ for thin resonators. When the membrane is subjected to temperature changes, the total tension is dominated by thermal strains $\epsilon^{th}_{r}$. Thermal strains are of dilatational nature and do not cause any shear, thus, these can be written as: $\epsilon^{th}_{r}=\alpha \Delta T$, where $\alpha$ is the thermal expansion coefficient. In-plane radial thermal strain is then related to tension according to the Hooke's law as $N=N_0 + Et\epsilon^{th}_r/(1-\nu)$, where $N_0$ the intrinsic pre-tension introduced during the fabrication process. Thus, equation (\ref{eq:SI_freq}) is used to determine if the membranes under study are in the plate or in the membrane limit at a given temperature, as shown in Fig. \ref{fgr:plate-membrane}a,b. 

The relative contribution of the plate term to the frequency of the resonator is largest at room temperature because the membrane tension is the lowest. As shown in Fig. \ref{fgr:plate-membrane}a at $T=300$ K and $N_0=0.1$ N/m, thicker FePS$_3$ samples ($t>40$ nm) behave as circular plates. However, as shown in Fig. \ref{fgr:plate-membrane}b, in proximity of the transition temperature ($T\approx114$ K) due to temperature-induced strain $\epsilon(T)$ (see equation (\ref{eq:SI_freq})), the resonator behaves close to the membrane limit over a thickness range from zero to $60$ nm. 

The total strain in the membrane is estimated using $\epsilon(T)=\epsilon_0-\int_{300K}^{T}(\alpha_{\mathrm{material}}(T)-\alpha_{\rm Si}(T))\mathrm{d}T$, where $\epsilon_0$ is the intrinsic pre-strain at $T=300$ K. Because $\alpha_{\mathrm{SiO_2}}\ll \alpha_{\mathrm{Si}}$ \cite{White1977,Lyon1977}, the effect of the thin SiO$_2$ layer can be neglected. As shown in Fig. \ref{fgr:plate-membrane}c, the thermal expansion coefficient of the silicon substrate (solid magenta line) is small compared to that of FePS$_3$. Therefore, the total strain in the membrane will mainly build up due to $\alpha_{\mathrm{FePS_3}}$, and this term dominates the change in resonance frequency, $f_0(T)$ as depicted in Fig. \ref{fgr:plate-membrane}d.

\section{Mechanical resonances and specific heat of MPS$_3$ (M=F\MakeLowercase{e}, N\MakeLowercase{i}, M\MakeLowercase{n})}

In addition to FePS$_3$, we measure MnPS$_3$ and NiPS$_3$ membranes with a $10$ $\mu$m diameter and thicknesses of $31.8\pm1.2$ and $35.7\pm1.1$ nm, respectively. Comparative study of these is particularly interesting since FePS$_3$ is an Ising antiferromagnet, while the other two are Heisenberg (MnPS$_3$) and XY (NiPS$_3$) antiferromagnets. The resonance peak of the fundamental membrane mode, $f_0(T)$, as well as the Q-factor is measured from $4-200$ K using the procedure described in the main text.\\
\begin{figure*}[ht]
	\includegraphics[scale=0.55]{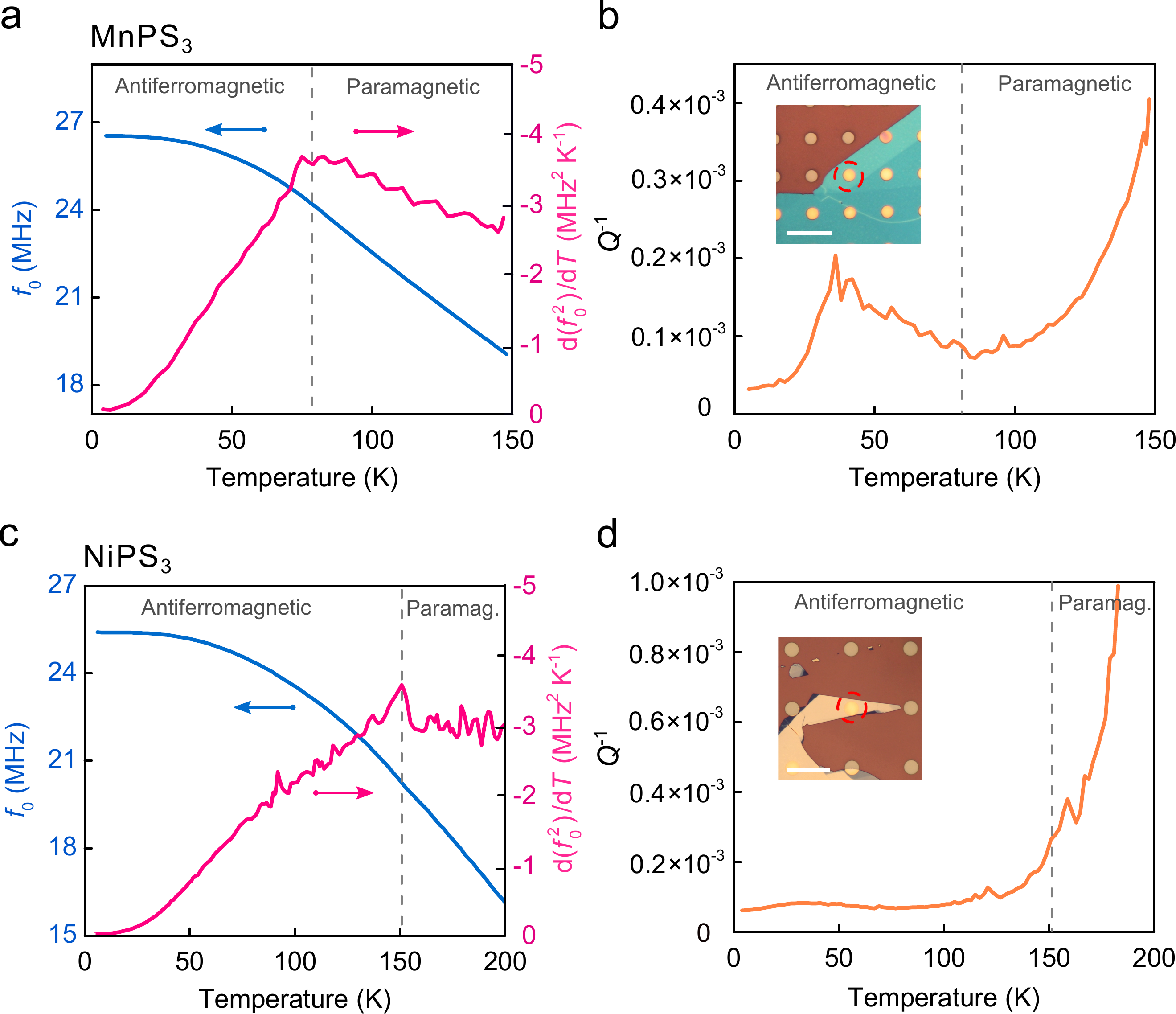}
	\caption{Mechanical properties of MnPS$_3$ and NiPS$_3$ membranes. Dashed vertical lines indicate transition temperatures, $T_N$. (a) Solid blue line - Measured resonance frequency of MnPS$_3$ membrane as a function of temperature. Solid magenta line - Temperature derivative of $f^2_0$. (b) Measured mechanical damping ($Q^{-1}$) as a function temperature. Inset: Optical image of the sample, $t=32.3\pm0.4$ nm. Scale bar: $40$ $\mu$m.
	(c) Solid blue line - Measured resonance frequency of NiPS$_3$ membrane as a function of temperature. Solid magenta line - Temperature derivative of $f^2_0$. (b) Measured mechanical damping ($Q^{-1}$) as a function temperature. Inset: Optical image of the sample, $t=35.7\pm0.5$ nm. Scale bar: $30$ $\mu$m.}
	\label{fgr:SI_MPS3}
\end{figure*}
At the phase transition, significant changes in both the resonance frequency (Fig. \ref{fgr:SI_MPS3}a) and quality factor (Fig. \ref{fgr:SI_MPS3}b) of the MnPS$_3$ membrane are observed. Figure \ref{fgr:SI_MPS3}a shows the resonance frequency (solid blue line) and the corresponding $\frac{{\rm d}(f_0^2)}{{\rm d}T}$ (solid magenta line) with a peak that occurs at a temperature similar to the transition temperature from the antiferromagnetic phase ($<78$ K) to the paramagnetic phase ($>78$ K) in the bulk material \cite{Takano2004} and is thus attributed as $T_N$. The temperature dependence of ${\rm d}(f^2_{0})/{\rm d}T$ shows a broad hump with a smeared peak at $T_N$, which resembles the temperature dependent specific heat $c_v(T)$ of this material in bulk form \cite{Takano2004}. Figure \ref{fgr:SI_MPS3}b shows the mechanical dissipation, $Q^{-1}$, that exhibits a local minimum close to $T_N$ as well as a local maximum at $T\sim36$ K. In Fig. \ref{fgr:SI_MPS3}c the resonance frequency of the NiPS$_3$ membrane (solid blue line) is shown with the corresponding $\frac{{\rm d}(f_0^2)}{{\rm d}T}$ (solid magenta line). A small peak is noticeable in $\frac{{\rm d}(f_0^2)}{{\rm d}T}$ near bulk $T_N\sim155$ K indicating the phase transition \cite{Kim2019}. However, as shown in Fig. \ref{fgr:SI_MPS3}d, no significant anomalies in the Q-factor were observed in the case of NiPS$_3$. Compared to FePS$_3$, the effect of the phase transitions in MnPS$_3$ and NiPS$_3$ on the resonances is more gradual (Fig. \ref{fgr:SI_MPS3}a-d). Both materials show a peak in $\frac{{\rm d}(f_0^2)}{{\rm d}T}$ at the $T_N$, but their dissipation does not show a clear jump at $T_N$ like in the case of FePS$_3$.  

\begin{figure*}[ht]
	\includegraphics[scale=0.24]{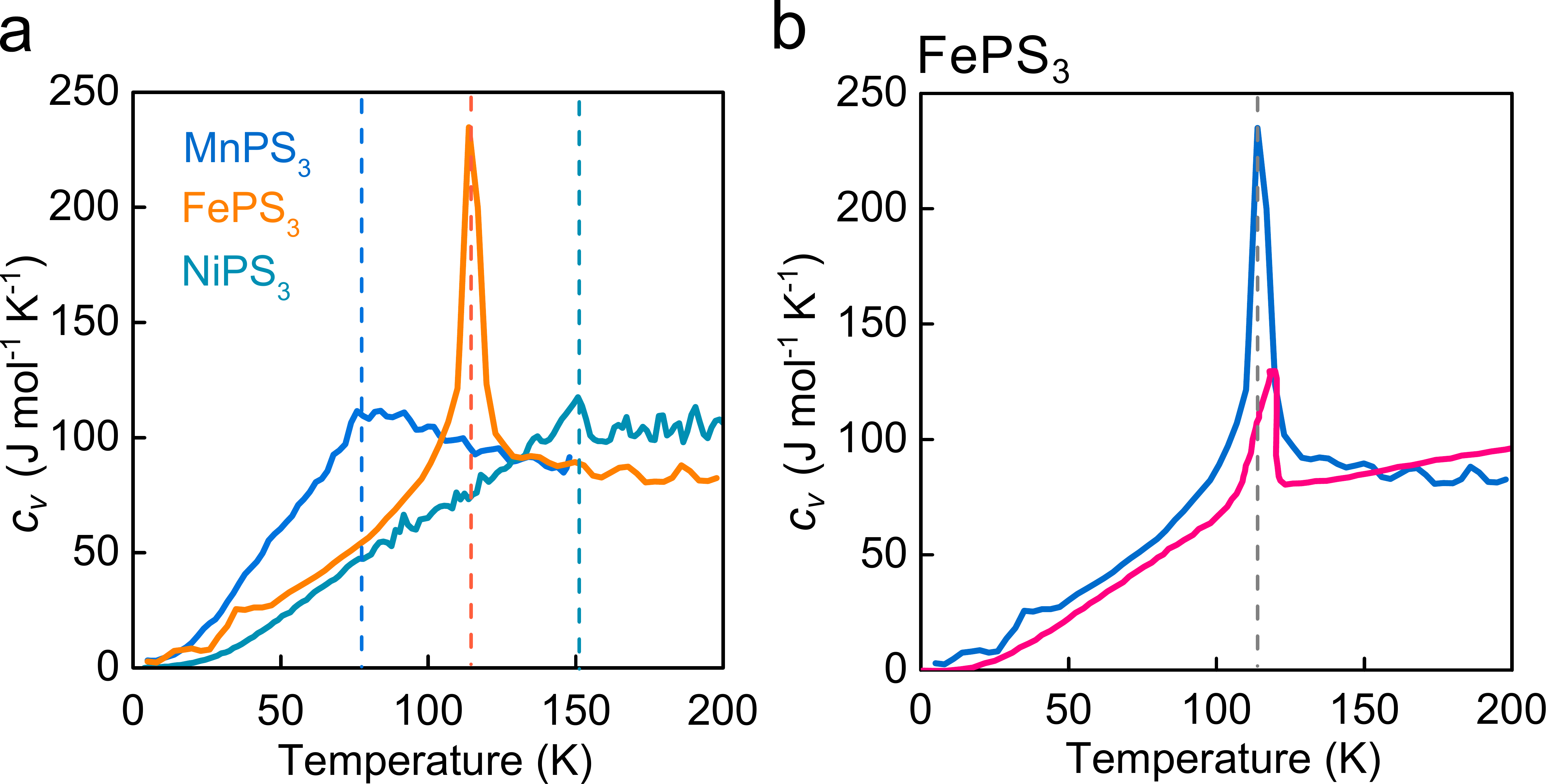}
	\caption{Estimated specific heat ($c_v$) for (a) MPS$_3$ (M=Fe, Mn, Ni) membranes of $t=45.2\pm0.6$ nm, $31.8\pm1.2$ nm and $35.7\pm1.1$ nm, respectively. Dashed lines indicate the corresponding transition temperatures ($T_N$): $T_N=76\pm5$ K for MnPS$_3$, $T_N=114\pm3$ K for FePS$_3$ and $T_N=151\pm5$ K for NiPS$_3$. (b) Specific heat of ultrathin FePS$_3$ membrane compared to that of its bulk form. Solid blue line - measured $c_v$ of the FePS$_3$ ultrathin membrane. Solid magenta line - $c_v$ reported by Takano, \textit{et al.}  \cite{Takano2004} for a bulk crystal.} 
	\label{fgr:SI_C_MPS3}
\end{figure*}

We also calculated the corresponding temperature dependent specific heat $c_v(T)$ for the three MPS$_3$ samples. Following the methodology described in the main text, we estimate the Gr\"{u}neisen parameter following the Belomestnykh$-$Tesleva relation \cite{Sanditov2009,Belomestnykh2004}: $\gamma\approx\frac{3}{2}\left(\frac{1+\nu}{2-3\nu}\right)$. We use reported values for monolayers of FePS$_3$ ($C_{11}=72.7$ N/m, $C_{12}=22.1$ N/m and $\rho_{\mathrm{2D}}=2.16\times 10^{-6}$ kg/m$^2$), MnPS$_3$ ($C_{11}=61.7$ N/m, $C_{12}=20.2$ N/m and $\rho_{\mathrm{2D}}=2.00\times 10^{-6}$ kg/m$^2$) and NiPS$_3$ ($C_{11}=87.0$ N/m, $C_{12}=23.1$ N/m and $\rho_{\mathrm{2D}}=2.15\times 10^{-6}$ kg/m$^2$) as obtained from first-principles calculations \cite{Hashemi2017,Zhang2016}. We convert these to three-dimentional Young's modulus $E$, mass density $\rho$ and Poisson's ratio $\nu$ using $E=\frac{C_{11}^2-C_{12}^2}{C_{11}}\frac{1}{t}$, $\rho=\frac{\rho_{\mathrm{2D}}}{t}$ and $\nu=\frac{C_{12}}{C_{11}}$, assuming the interlayer distance for the compounds to be $t_{\mathrm{FePS_3}}=0.64$ nm, $t_{\mathrm{MnPS_3}}=0.65$ nm and $t_{\mathrm{NiPS_3}}=0.64$ nm as determined from Fig. \ref{fgr:SI_crystals}a-c. The resulting values are $E=103$ GPa, $\nu=0.304$ and $\rho=3375$ kg/m$^3$ for FePS$_3$, $E=85$ GPa, $\nu=0.327$ and $\rho=3076$ kg/m$^3$ for MnPS$_3$ and $E=126$ GPa, $\nu=0.265$ and $\rho=3359$ kg/m$^3$ for NiPS$_3$. 

In Fig. \ref{fgr:SI_C_MPS3}a the specific heat for the three MPS$_3$ are displayed, as determined from the data in Fig. \ref{fgr:SI_MPS3}a,c, and Fig. 2a using equation (2) from the main text. Detected temperatures are indicated by dashed lines and found to be $T_N=76\pm5$ K for MnPS$_3$, $T_N=114\pm3$ K for FePS$_3$ and $T_N=151\pm5$ K for NiPS$_3$, which are in agreement with experiments in bulk crystals of MPS$_3$ \cite{Takano2004, Joy1992}. As shown in Fig. \ref{fgr:SI_C_MPS3}b, for FePS$_3$ we obtain a good correspondence to bulk literature values \cite{Takano2004} (solid magenta curve) without fitting parameters. The transition related peak in $c_v$, however, is more pronounced in the case of the ultrathin membrane (solid blue curve) than in the bulk.

\section{Entropy in a suspended antiferromagnet}
A uniaxial antiferromagnet with two antiparallel magnetic sub-lattices close to the phase transition can be modeled by the Landau theory of phase transitions \cite{Landau:1937obd,Landau1984}. In this theory, the vector order parameter $\boldsymbol{L}$ is defined as the difference between the magnetizations of the two sublattices, $\boldsymbol{M_1}$ and $\boldsymbol{M_2}$ (thus, the antiferromagnetic vector $\boldsymbol{L}=\boldsymbol{M_1}-\boldsymbol{M_2}$). This order parameter is zero in the paramagnetic phase and is finite in the antiferromagnetic phase. The magnetization, $\boldsymbol{M}$, is the sum of both magnetizations ($\boldsymbol{M}=\boldsymbol{M_1}+\boldsymbol{M_2}$) and equals to zero in the absence of an external magnetic field, $\boldsymbol{H}$.

For a uniaxial crystal antiferromagnet, the transition temperature is commonly known as N\'{e}el temperature, $T_N$, but, as conveyed by Landau, \textit{et al.} \cite{Landau:1937obd,Landau1984}, it is also referred to as the antiferromagnetic Curie temperature and denoted as $T_c$. Note that further we will consider the N\'{e}el temperature, denoted as $T_N$, as the temperature where the transition from a paramagnetic to an antiferromagnetic phase takes place. Near $T_N$, $\boldsymbol{L}$ is small and the free energy, $F$, can be expanded in terms of $\boldsymbol{L}$ and $\boldsymbol{H}$, since the magnetization is only non-zero when an external field $\boldsymbol{H}$ is present (i.e. a spin-flop transition). Following Landau formalism and considering $z$-axis as the main axis of symmetry, we write (see e.g. Ref. \onlinecite{Landau1984}):
\begin{equation} \label{Landau_AFM_no_strain}
F = F_0 + AL^2 + BL^4 + D \left(\boldsymbol{H} \cdot \boldsymbol{L}\right)^2 + D' H^2 L^2 - \frac{1}{2} \chi_p H^2 + \frac{1}{2} \beta \left( L_x^2 + L_y^2 \right) - \frac{1}{2} \mu_M \left( H_x^2 + H_y^2 \right)-\frac{H^2}{8\pi}\ , 
\end{equation}
where $A = a (T - T_N)$, $D$, $D'$, $a$ and $B$ are phenomenological positive constants which are taken temperature independent, $\chi_p$ the isotropic susceptibility for $T > T_N$, $\beta$ the index that describes the temperature dependence of the spontaneous magnetization below $T_N$ ($\beta > 0$ for $\boldsymbol{L}$ directed out-of-plane) and $\mu_M$ the magnetic susceptibility in the paramagnetic phase. The minimization of equation (\ref{Landau_AFM_no_strain}), where the vector $\boldsymbol{L}$ is along the $z$-axis (thus, $L_x = L_y = 0$ and $\beta > 0$) and in the absence of field ($H = 0$), gives $L_x = L_y = 0$ and $L_z = 0$ for $T>T_N$ (paramagnetic phase) and $L_z = \sqrt{a (T_N - T)/(2B)}$ for $T < T_N$ (antiferromagnetic phase).

Now we introduce strain. For an easy-axis antiferromagnet near $T_N$ with the vector $\boldsymbol{L}$ along the $z$-axis and in the absence of field, we can write equation (\ref{Landau_AFM_no_strain}) as:
\begin{equation} \label{Landau_AFM}
F = F_0 + AL^2 + BL^4 + \frac{1}{2} \beta \left( L_x^2 + L_y^2 \right) + \zeta L_z^2 + \zeta_x L_x^2 + \zeta_y L_y^2 \ , 
\end{equation}
where the last three added terms describe the magnetostriction effects, {i.e.}, the coupling of magnetic moments to strain; the coefficients $\zeta$, $\zeta_{x,y}$ are linear combinations of the components of the strain tensor. We assume that the strain is determined by the deformation of the membrane, and that the back-action exerted by the magnetization on the strain is negligible. In that case, for the calculation of the order parameter, $\zeta$ and $\zeta_{x,y}$ can be treated as temperature-independent constants. We also assume that $\vert \zeta \vert, \vert \zeta_{x,y} \vert \ll \beta$, so that even the strained antiferromagnet exhibits an easy-axis. The minimization of equation (\ref{Landau_AFM}) gives $L_x = L_y = 0$, leading to a free energy which only depends on $L_z$,
\begin{equation} \label{Landau_AFM_strain}
F = F_0 + \left[ a \left(T - T_N \right) + \zeta \right] L_z^2 + BL_z^4. 
\end{equation}
The first observation is that the magnetostriction effects shift the antiferromagnetic phase transition point. Indeed, the phase transition occurs at the temperature $T_N^*$ at which the coefficient multiplying $L_z^2$ in equation (\ref{Landau_AFM_strain}) vanishes. This gives $T_N^* = T_N - \zeta/a$ (which is equation ($4$) in the main text).

Second, we calculate the behavior of the specific heat close to the phase transition. Minimizing equation (\ref{Landau_AFM_strain}) with respect to $L_z$, we find the equilibrium free energy, $F_{min} = F_0 -a^2 (T - T_N^*)^2/(4B)$ in the antiferromagnetic phase, where $F_0$ is the free energy of the paramagnetic phase. We proceed by calculating the entropy $S_{min} = -\partial F_{min} / \partial T$,
\begin{eqnarray} \label{entropy_no_strain}
S_{min} - S_0 = \left\{ \begin{array}{lr} - a^2 (T_N^* - T)/(2B) & T < T_N^* \\ 0 & T > T_N^* \end{array} \right.  \ ,
\end{eqnarray}
where $S_0 = -\partial F_0/\partial T$ is the entropy of the paramagnet with $S_0$ the non-magnetic contribution to the entropy. Since the renormalized transition temperature $T_N^*$ is strain-dependent, the entropy of the antiferromagnet contains an additional (as compared to the paramagnetic phase) strain-dependent term. 
The specific heat near $T_N^*$, $c_{v,min} = T \partial S_{min}/\partial T$, reads 
\begin{eqnarray} \label{specific_heat_strain}
c_{v,min} - c_{v,0} = \left\{ \begin{array}{lr} T_N^* a^2/(2B) & T < T_N^* \\ 0 & T > T_N^* \end{array}  \right.  \ ,
\end{eqnarray}
where $c_{v,0}$ is the non-magnetic contribution to the specific heat. This derivation shows, in line with Landau theory \cite{Landau:1937obd,Landau1984}, that the specific heat has a jump at the transition temperature, $T_N^*$.

\section{Dissipation and thermoelastic damping in vibrating membranes}

For a membrane in motion the dissipation $Q^{-1}$ is defined as the ratio of the energy lost per cycle to $2 \pi$ times the stored energy. The total dissipation in a membrane is given by the sum of all contributing dissipation mechanisms \cite{Schmid2016}:
\begin{equation} \label{LifshitzRoukes_Q_general}
Q^{-1}=Q^{-1}_{medium}+Q^{-1}_{clamping}+Q^{-1}_{intrinsic}+Q^{-1}_{other}.
\end{equation}
$Q^{-1}_{medium}$ is related to losses due to the interaction with a fluid medium or gas and can thus be neglected in high vacuum. $Q^{-1}_{clamping}$ is related to the transfer of mechanical energy to the anchoring substrate, which has a small temperature dependence \cite{Schmid2016}. $Q^{-1}_{intrinsic}$ quantifies all intrinsic damping mechanisms of the material, such as thermoelastic damping, internal and surface friction, and phonon-phonon interaction loss. $Q^{-1}_{other}$ relates all other possible damping mechanisms, such as electrical charge damping and magneto-motive damping. In this section we focus on deriving the expression for the thermoelastic damping. Further derivation in this section follows Zener's standard linear solid model \cite{Zener1937,Zener1938}.

The dissipation is equal to the ratio between the imaginary and real parts of the complex elastic modulus $E^{*}(\omega)=E'(\omega)+i E''(\omega)$:
\begin{equation}\label{eq:Lifshitz_e_e}
Q^{-1}=\frac{E''}{E'}.
\end{equation}
For a standard linear solid with a single relaxation mechanism, real and imaginary parts of the complex elastic modulus are given by:
\begin{equation} 
E{'}(\omega)=E_r+(E_u-E_r)\frac{\omega^2 \tau^2}{1+(\omega \tau)^2},\:
E{''}(\omega)=(E_u-E_r)\frac{\omega \tau}{1+(\omega \tau)^2},
\end{equation} 
where $\omega$ is the resonance eigenfrequency, $\tau$ the thermal relaxation time, $E_r$ and $E_u$ are the relaxed (or isothermal) and unrelaxed (or adiabatic) Young's moduli, respectively (see Ref. \onlinecite{Schmid2016,Zener1938} for more details). From equation (\ref{eq:Lifshitz_e_e}), the dissipation is then:
\begin{equation} \label{eq:LifshitzRoukes_Q}
Q^{-1}=(E_u-E_r) \frac{\omega \tau}{E_r+E_u(\omega \tau)^2}\approx\frac{E_u-E_r}{E_r} \frac{\omega \tau}{1+(\omega \tau)^2},
\end{equation}
for $(E_u-E_r)\ll E_r\approx E_u$. For a standard thermoelastic solid and in the case of thermoelastic damping $Q^{-1}_{TED}$, equation (\ref{eq:LifshitzRoukes_Q}) can be rewritten as \cite{Zener1938, Zener1937, Lifshitz2000}:
\begin{equation} \label{eq:LifshitzRoukes_Q_first}
Q^{-1}_{TED}=\frac{E_u-E_r}{E_r}\beta=\frac{E\alpha^2T}{c_v}\beta,
\end{equation}
where $\alpha$ is the thermal expansion coefficient, $c_v$ the specific heat and $\beta$ the thermomechanical parameter, that in Zener's model is $\beta_{Z} = \frac{\omega \tau}{1+(\omega \tau)^2}$. The exact expression for thermoelastic damping and its relation to the thermal properties of solids was found by Lifshitz and Roukes \cite{Lifshitz2000} with $\beta_{LR} = \frac{6}{\xi^2}-\frac{6}{\xi^3}\frac{\sinh(\xi)+\sin(\xi)}{\cosh(\xi)+\cos(\xi)}$, where $\xi=\frac{\pi}{\sqrt{2}}\sqrt{\omega \tau}$. Under the assumption that the temperature dependence of $\omega \tau$ is small, equation (\ref{eq:LifshitzRoukes_Q_first}) can be written as:
\begin{equation} \label{eq:LifshitzRoukes_Q_general}
Q^{-1}_{TED}\propto \frac{E\alpha^2T}{c_v}.
\end{equation}
The close correspondence between this expression and the data in Fig. 2d and 4d in the main text indicates that this assumption is reasonable.\\
As could be noted from equation (2) in the main text and assuming the elastic properties of the material and its Gr\"{u}neisen parameter to have a negligible temperature dependence at low temperatures, the thermoelastic damping $Q^{-1}_{TED}$ is related to the frequency $f_0$ of the resonator as:
\begin{equation} 
Q^{-1}_{TED}(T)\propto \frac{E\alpha^2(T)T}{c_v(T)}\propto c_v(T)\, T \propto - T\frac{{\rm d}[f_0^2(T)]}{{\rm d}T},
\end{equation}
where we use $\alpha (T) \propto c_v(T)$ and $c_v(T) \propto \frac{{\rm d}[f_0^2(T)]}{{\rm d}T}$ according to equation (2) in the main text. Therefore, in accordance with our observations in the main text, an anomaly (a jump) in the specific heat at the transition temperature will be visible in both $\frac{{\rm d}[f_0^2(T)]}{{\rm d}T}$ and $Q^{-1}(T)$ if the thermoelastic damping is the dominating dissipation mechanism.

\section{Electric field induced strain in a circular F\MakeLowercase{e}PS$_3$ membrane}

\begin{figure*}[ht]
	\includegraphics[scale=0.65]{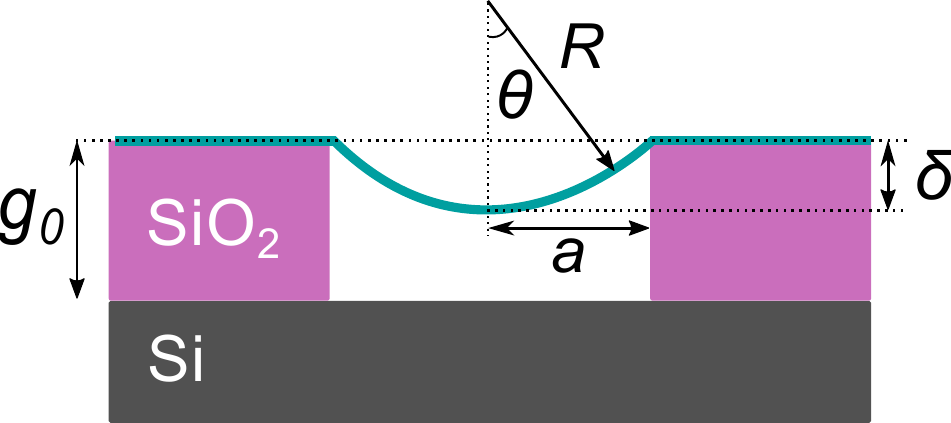}
	\caption{Schematic of the membrane deformation, cross-section view.}
	\label{fgr:SI_scheme_strain}
\end{figure*}

\begin{figure*}[ht]
	\includegraphics[scale=0.65]{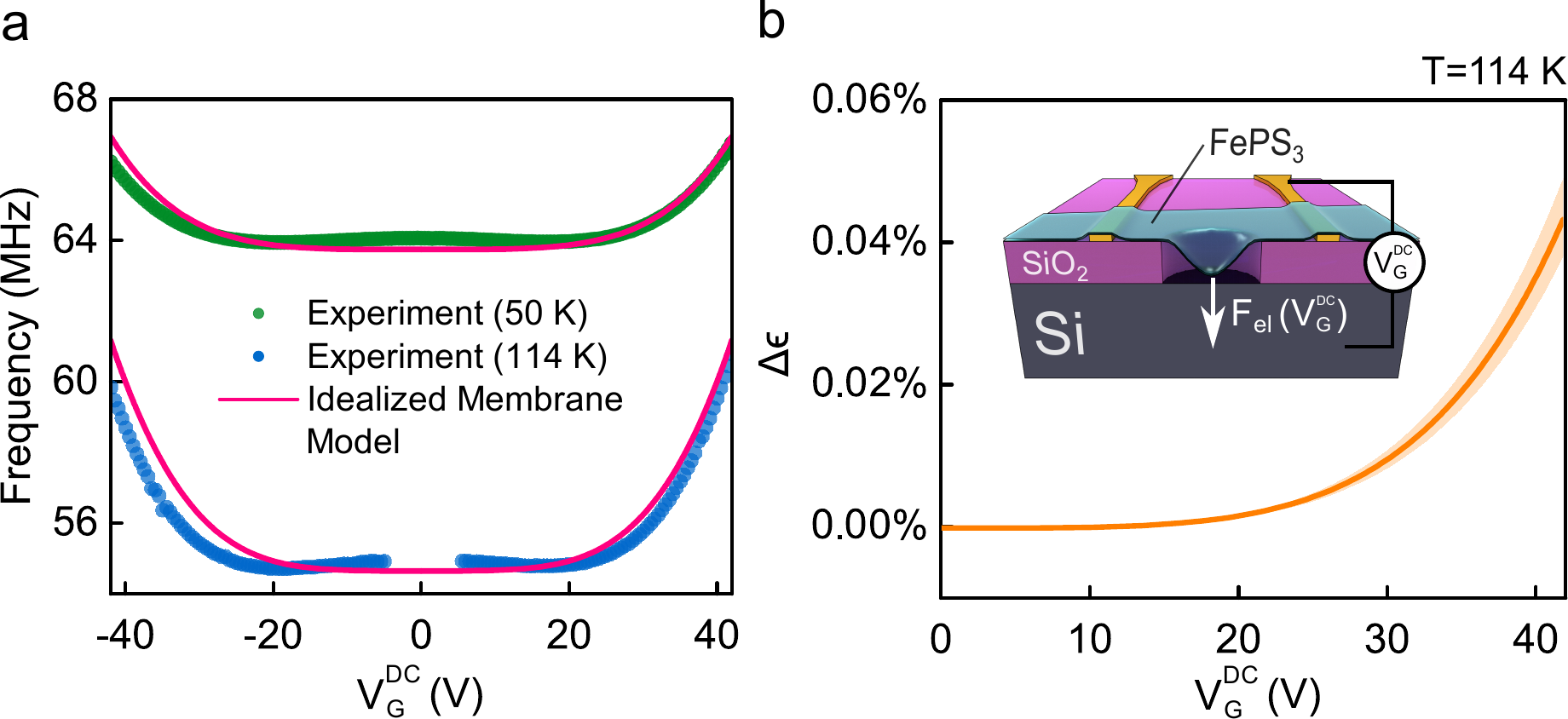}
	\caption{(a) Idealized electrostatically strained membrane model for $N_0=0.01$ N/m, $g_0=285$ nm, $a=2$ $\mu$m, $t=8$ nm, $E=103$ GPa and $\nu=0.304$; superimposed with corresponding experimental data. (b) Estimate of added radial strain at $T=114$ K. The shaded orange region represents an estimated uncertainty from the accuracy in determining the temperature induced strain ($\pm0.015\%$) at $T=114$ K. Inset - schematics of the experiment.}
	\label{fgr:SI_strain_fit}
\end{figure*}

A constant electrostatic load is applied to the circular membrane with a radius $a$ (see Fig. \ref{fgr:SI_scheme_strain}). This results in a uniform curvature $R$ with a maximum deflection $\delta$. 
\\
In this system, the electrostatic load is balanced by the total tension $N_{\mathrm{total}}$:
\begin{equation}
    \frac{\varepsilon_0 (V_G^{\mathrm{DC}})^2}{2(g_0-\delta)^2}\pi R^2=2\pi N_{\mathrm{total}} R,
\end{equation}
where $\varepsilon_0$ is the vacuum permittivity, $V_G^{\mathrm{DC}}$ the applied voltage, $g_0$ the gap size between the membrane and the bottom silicon plate. For small deflections the radius of curvature can be approximated as:
\begin{equation}\label{eq:strain_radius}
    R\approx \frac{a^2}{2\delta}, 
\end{equation}
so that the tension $N_{\mathrm{total}}$ becomes:
\begin{equation}\label{eq:strain_radius_tension}
    N_{\mathrm{total}}=\frac{\varepsilon_0(V_G^{\mathrm{DC}})^2}{2(g_0-\delta)^2}\frac{a^2}{4\delta}.
\end{equation}
The radial strain in such a membrane can be estimated from the arc length \cite{Small1992, Weber2014}:
\begin{equation}
    \epsilon=\frac{R\theta-a}{a}\approx \frac{a^2}{6R^2}.
\end{equation}
Combining this result with equation (\ref{eq:strain_radius}) yields:
\begin{equation}\label{eq:strain_deflection}
    \epsilon \approx \frac{2\delta^2}{3a^2}.
\end{equation}
From Hooke's law and including equation (\ref{eq:strain_deflection}), one can write the strain due to deformation as:
\begin{equation}
    N=\frac{Et}{1-\nu}\epsilon=\frac{2Et\delta^2}{3a^2(1-\nu)},
\end{equation}
Therefore, the total tension in the membrane, including the thermal expansion induced tension $N_0(T)$ at a certain temperature $T$, can be written as:
\begin{equation}
N_{\mathrm{total}}=N_0(T)+N=N_0(T)+\frac{2Et\delta^2}{3a^2(1-\nu)}=\frac{\varepsilon_0(V_G^{\mathrm{DC}})^2}{2(g_0-\delta)^2}\frac{a^2}{4\delta},
\end{equation}
which also can be rewritten as: 
\begin{equation}\label{eq:SI_strain_voltage}
V_G^{\mathrm{DC}}=\sqrt{\frac{2(g_0-\delta)^2}{\varepsilon_0}\left(\frac{4\delta N_0(T)}{a^2}+\frac{8Et\delta^3}{3a^4(1-\nu)}\right)}.
\end{equation}
As shown in Fig. \ref{fgr:SI_strain_fit}a, equation (\ref{eq:SI_strain_voltage}) in combination with (\ref{eq:strain_deflection}) and (\ref{eq:SI_freq}) fits the experimental data well and is used to provide an an estimate of the electrostatically induced strain in the membrane (see Fig. \ref{fgr:SI_strain_fit}b). 

\section{Mechanical resonances and specific heat of 2H-T\MakeLowercase{a}S$_2$ near the charge density wave transition} 

\begin{figure*}[ht]
	\includegraphics[scale=0.63]{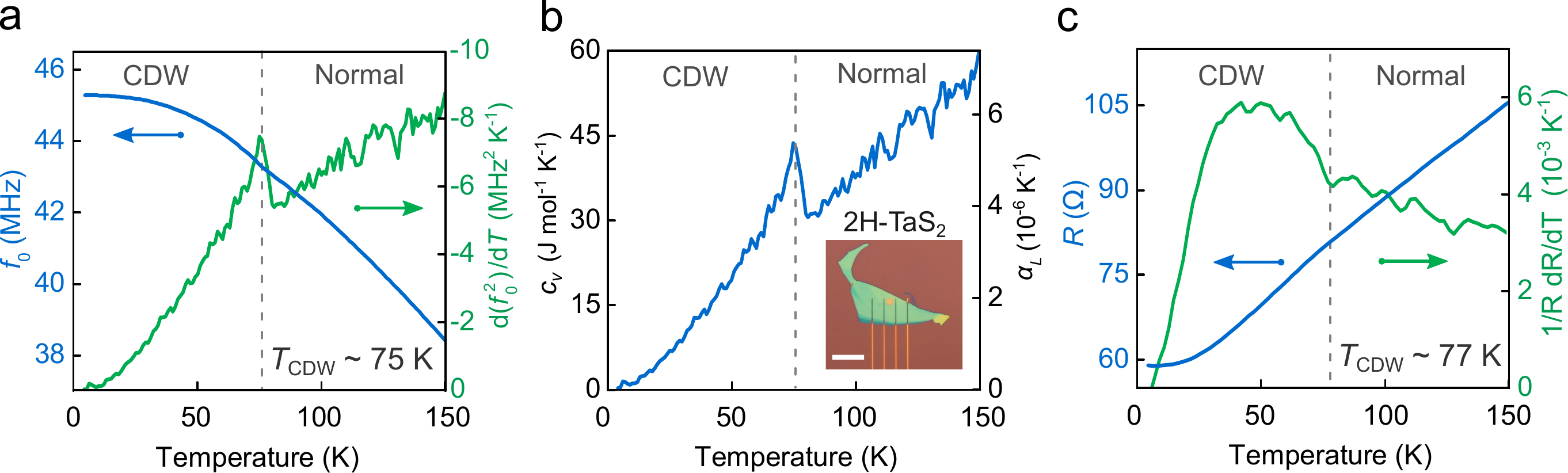}
	\caption{Mechanical properties of a 2H-TaS$_2$ resonator with membrane thickness of $31.2\pm0.6$ nm. Dashed vertical lines in the panels indicate the transition temperature, $T_{\mathrm{CDW}}$. (a) Solid blue line - resonance frequency as a function of temperature. Solid green line - temperature derivative of $f^2_0$. (b) Solid blue line - estimated specific heat ($c_v$) and thermal expansion coefficient ($\alpha_L$). Inset: optical image of the sample. Scale bar: $20$ $\mu$m. (c) Solid blue line - four-point resistance of the same sample. Solid green line - $\frac{1}{R}\frac{dR}{dT}$ plot showing the CDW related feature at $T_{\mathrm{CDW}}$. }
	\label{fgr:SI_Tas2}
\end{figure*}

For 2H-TaS$_2$ resonance frequency measurements were performed on a $d=4$ $\mu$m drum made of a $31$ nm thin flake (see Fig. \ref{fgr:SI_Tas2}a, solid blue line). As shown by the solid green line in Fig. \ref{fgr:SI_Tas2}a, the specific heat-related temperature derivative of $f^2_0$ reveals a clear peak at $T_{CDW}\sim75$ K. We convert the measured $\frac{{\rm d}(f_0^2)}{{\rm d}T}$ to the specific heat using the methodology described in the main text. We use reported values of $E_{\mathrm{2D}}=87$ N/m and $\nu=0.27$ for a monolayer of 1H-TaS$_2$ obtained from molecular dynamics simulations \cite{jiang2017parameterization} and mass density $\rho=6110$ kg/m$^3$. We find the corresponding $E=E_{\mathrm{2D}}/t=149$ GPa, taking the interlayer spacing $t=0.58$ nm. The estimated specific heat of the 2H-TaS$_2$ membrane is depicted in Fig. \ref{fgr:SI_Tas2}b (solid blue line). The four-probe resistance was measured on the same flake to confirm the existence of a CDW transition using a conventional electronic based method as shown in Fig. \ref{fgr:SI_Tas2}c. The expected characteristic kink in the resistance is visible at $\sim77$ K, consistent with the CDW transition temperature previously reported in 2H-TaS$_2$ \cite{AbdelHafiez2016}.

\section{Crystal growth and characterization}

The crystal growth of MPS$_3$ (M = Mn, Fe, Ni) was performed following a typical solid state reaction. Powders of Mn ($>99.9\%$, from Sigma-Aldrich), Fe ($99.998\%$, from Alfa-Aesar), Ni ($99.99\%$, from Sigma-Aldrich), P ($> 99.99 \%$, from Sigma-Aldrich) and S ($99.998\%$, from Sigma-Aldrich) were mixed in a stoichiometric ratio, pressed into a pellet and sealed in an evacuated quartz ampoule ($P \sim5\times10^{-5}$ mbar, length $= 25$ cm, internal diameter $= 1.5$ cm) and heated from room temperature to $400$ $^{\circ}$C at $1.1$ $^{\circ}$C/min. Then, the temperature was kept constant for twenty days and slowly cooled down ($0.07$ $^{\circ}$C/min).

\begin{figure*}[ht]
	\includegraphics[scale=1]{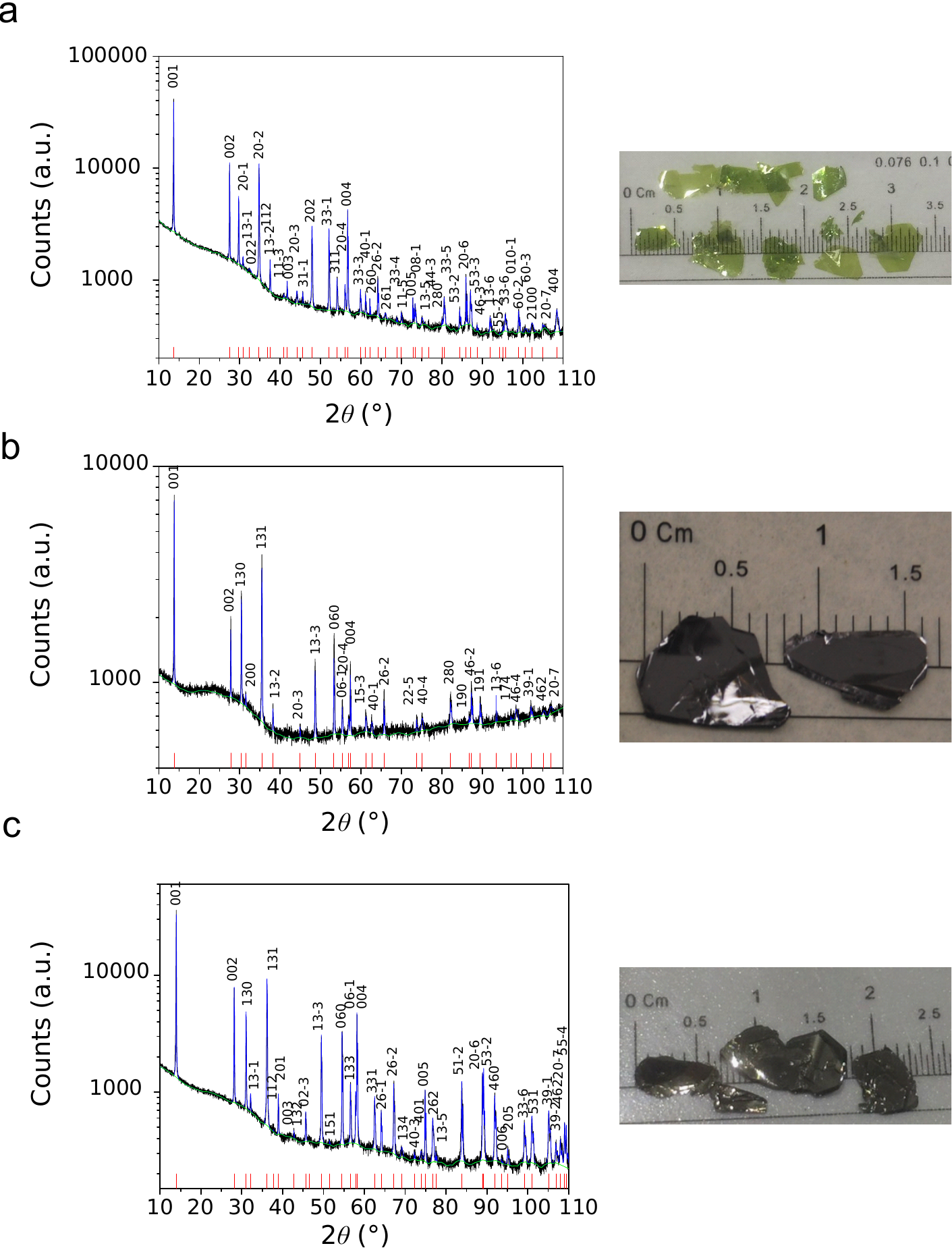}
	\caption{XRPD experimental patterns (black) and corresponding fits (peaks in blue and background in green) for MPS$_3$ (M = Mn, Fe and Ni) crystals. For clarity, the peaks have been marked by red lines. (a) Left: Pattern of MnPS$_3$. Right: Photography of a MnPS$_3$ crystal. The fit gives as a result: a = 6.077(7) {\AA}, b = 10.55(2) {\AA} and c = 6.805(9) {\AA}, $\alpha = \gamma = 90^{\circ}$ and $\beta = 107.33(1)^{\circ}$,  monoclinic C face center crystal system with C12/m1 space group, $\chi^2 = 5.8 \times 10^{-5}$.  (b) Left: Pattern of FePS$_3$. Right: Photography of a FePS$_3$ crystal. The fit gives as a result: $a = 5.939(6)$ {\AA}, $b = 10.296(3)$ {\AA} and $c = 6.716(3)$ {\AA}, $\alpha = \gamma = 90^{\circ}$ and $\beta = 107.13(1)^{\circ}$,  monoclinic C face center crystal system with C12/m1 space group, $\chi^2 = 1.5 \times 10^{-6}$. (c) Left: Pattern of NiPS$_3$. Right: Photography of a NiPS$_3$ crystal. The fit gives as a result: $a = 5.815(4)$ {\AA}, $b = 10.087(5)$ {\AA} and $c = 6.627(4)$ {\AA}, $\alpha = \gamma = 90^{\circ}$ and $\beta = 106.945(9)^{\circ}$,  monoclinic C face center crystal system with C12/m1 space group, $\chi^2 = 1.1 \times 10^{-5}$.}
	\label{fgr:SI_crystals}
\end{figure*}

For obtaining large crystals, $4$ mmol of the previous material was mixed with I$_2$ as a transport agent ([I$_2$] $\sim5$ mg/cm$^3$) in an evacuated quartz ampoule ($P \sim5\times10^{-5}$ mbar, length $= 50$ cm, internal diameter $= 1.5$ cm). The quartz tube was placed inside a three-zone furnace with the material in the leftmost zone. The other two zones were heated up in $24$ h from room temperature to $650$ $^{\circ}$C and kept at that temperature for one day. After this, the leftmost side was heated up to $700$ $^{\circ}$C in $3$ h and a gradient of $700$ $^{\circ}$C/$650$ $^{\circ}$C/$675$ $^{\circ}$C was established in the three-zone furnace. Then the temperature was kept constant for $28$ days and cooled down naturally. As shown in Fig. \ref{fgr:SI_crystals}, with this process we could obtain crystals with a length up to several centimeters. The obtained crystals were analyzed by ICP-OES (Inductively Coupled Plasma - Optical Emission Spectrometry) and powder X-ray diffraction. The relative weights of elements obtained are summarized in Table S\ref{tbl:SI_crystals}. The refinement of the X-ray diffraction pattern (Fig. \ref{fgr:SI_crystals}) revealed a monoclinic base-centered crystal system with C12/m1 space group and a unit cell determined by $\alpha = \gamma = 90^{\circ}$ and $\beta = 107.33(1)^{\circ}$ (MnPS$_3$), $\beta = 107.13(1)^{\circ}$ (FePS$_3$), $\beta = 106.945(9)^{\circ}$ (NiPS$_3$),  and $a = 6.077(7)$ {\AA}, $b = 10.55(2)$ {\AA}  and $c = 6.805(9)$ {\AA} for MnPS$_3$, $a = 5.939(6)$ {\AA}, $b = 10.296(3)$ {\AA} and $c = 6.716(3)$ {\AA} for FePS$_3$ and $a = 5.815(4)$ {\AA}, $b = 10.087(5)$ {\AA} and $c = 6.627(4)$ {\AA} for NiPS$_3$. The obtained results are in accordance with the ones reported in the literature \cite{Ouvrard1985}.

The crystal growth and characterization of 2H-TaS$_2$ was performed as already reported in earlier works \cite{PinillaCienfuegos2016,NavarroMoratalla2016}.

\begin{table*}[ht]
\label{tbl:SI_crystals}
\centering
\begin{tabular}{|l|l|l|l|}
\hline
\multicolumn{2}{|l|}{\textbf{Element}} & \textbf{\begin{tabular}[c]{@{}l@{}}Obtained\\ (\%)\end{tabular}} & \textbf{\begin{tabular}[c]{@{}l@{}}Expected\\ (\%)\end{tabular}} \\ \hline
\multirow{3}{*}{\textbf{MnPS$_3$}} & \textbf{Mn} & $28.0 \pm 1.0$ & $30.2$ \\ \cline{2-4} 
 & \textbf{P} & $12 \pm 1$ & $17.0$ \\ \cline{2-4} 
 & \textbf{S} & $49 \pm 2$ & $52.8$ \\ \hline
\multirow{3}{*}{\textbf{FePS$_3$}} & \textbf{Fe} & $30.0 \pm 1.0$ & $30.5$ \\ \cline{2-4} 
 & \textbf{P} & $15.7 \pm 0.5$ & $16.9$ \\ \cline{2-4} 
 & \textbf{S} & $51 \pm 2$ & $52.6$ \\ \hline
\multirow{3}{*}{\textbf{NiPS$_3$}} & \textbf{Ni} & $30.0 \pm 1.0$ & $31.6$ \\ \cline{2-4} 
 & \textbf{P} & $15.7 \pm 0.5$ & $16.6$ \\ \cline{2-4} 
 & \textbf{S} & $53 \pm 2$ & $51.8$ \\ \hline
\end{tabular}%
\caption{Experimental and expected relative weights analyzed by inductively coupled plasma - optical emission spectrometry (ICP-OES) for the different MPS$_3$ crystals (M = M\MakeLowercase{n}, F\MakeLowercase{e} and N\MakeLowercase{i}).}

\end{table*} 
\appendix